\documentclass[preprint,showpacs,preprintnumbers,amsmath,amssymb,floatfix]{revtex4}
\usepackage{epsfig}
\usepackage{wrapfig}
\begin{document}

\title{
\vskip-3cm{\baselineskip14pt
\centerline{\normalsize\rm DESY 04-181\hfill ISSN 0418-9833}
\centerline{\normalsize\rm MPP-2004-116\hfill}
\centerline{\normalsize\rm hep-ph/0409303\hfill}
\centerline{\normalsize\rm September 2004\hfill}}
\boldmath
Next-to-leading-order predictions for $D^{*\pm}$ plus jet photoproduction at
DESY HERA
\unboldmath}

\author{Gudrun Heinrich}

\affiliation{II. Institut f\"ur Theoretische Physik, Universit\"at Hamburg,
Luruper Chaussee 149, 22761 Hamburg, Germany}

\author{Bernd A. Kniehl}\thanks{Permanent address:
II. Institut f\"ur Theoretische Physik, Universit\"at Hamburg,
Luruper Chaussee 149, 22761 Hamburg, Germany}

\affiliation{Max-Planck-Institut f\"ur Physik (Werner-Heisenberg-Institut),
F\"ohringer Ring 6, 80805 Munich, Germany}

\thispagestyle{empty}

\begin{abstract}
We study the photoproduction of a $D^{*\pm}$ meson in association with a
hadron jet at next-to-leading order in the parton model of QCD with
non-perturbative fragmentation functions extracted from LEP1 data of $e^+e^-$
annihilation.
The transverse-momentum and rapidity distributions recently measured at DESY
HERA in various kinematic ranges nicely agree with our theoretical
predictions.
This provides a useful test of the universality and the scaling violations of
the fragmentation functions predicted by the factorization theorem.
These comparisons also illustrate the significance of the charm component in 
the resolved photon.
This is elaborated by investigating the cross-section distributions in
$x_{\rm obs}^\gamma$ and $\cos\theta^*$. 
\end{abstract}


\maketitle

\section{Introduction}

Heavy-flavor production has always been an important testing ground for
quantum chromodynamics (QCD), one of the reasons being that it addresses the
problem of where to draw the dividing line between perturbative and
non-perturbative aspects. 
The photoproduction of open charm is particularly interesting because it also
allows valuable insights into the partonic structure of the photon.
Results on inclusive $D^{*\pm}$ photoproduction from the H1
\cite{Adloff:1998vb} and ZEUS \cite{Breitweg:1997gc,Breitweg:1998yt}
Collaborations at the DESY $ep$ collider HERA have been compared to
leading-order (LO) and next-to-leading order (NLO) calculations
\cite{fo,bkk,bkkc}.
More recently, data on the photoproduction of a $D^{*\pm}$ meson in
association with a hadron jet have also become available
\cite{Breitweg:1998yt,Chekanov:2003bu,kohno,zeus,h1}.

Concerning the theoretical treatment of open heavy-flavor production, several 
approaches have been followed in the literature. 
The QCD-improved parton model implemented in the modified minimal-subtraction
($\overline{\rm MS}$) renormalization and factorization scheme and endowed
with non-perturbative fragmentation functions (FFs), which proved itself so
convincingly for light-hadron inclusive production \cite{kkp}, also provides
an ideal theoretical framework for a coherent global analysis of $D$-
\cite{bkkc} and $B$-meson data \cite{bkkb}, provided that $\mu\gg m_Q$, where
$\mu$ is the energy scale characteristic for the respective production process
and $Q=c,b$.
Then, at LO (NLO), the dominant logarithmic terms, of the form
$\alpha_s^n\ln^n\left(\mu^2/m_Q^2\right)$
($\alpha_s^{n+1}\ln^n\left(\mu^2/m_Q^2\right)$) with $n=1,2,\ldots$, where
$\alpha_s$ is the strong-coupling constant, are properly resummed to all
orders by the time-like Dokshitzer-Gribov-Lipatov-Altarelli-Parisi (DGLAP)
\cite{dglap} evolution, while power terms of the form
$\left(m_Q^2/\mu^2\right)^n$ are negligibly small and can be safely neglected.
In this {\it massless-quark scheme} or {\it zero-mass variable-flavour-number
scheme} (ZMVFNS), which is sometimes improperly referred to as {\it NLL
approximation} \footnote{%
The non-logarithmic corrections of relative order $\alpha_s$ are fully
included, except for terms that are suppressed by powers of
$\left(m_Q^2/\mu^2\right)^n$.},
the $Q$ quark is treated as massless and appears as an active parton in the
incoming hadron or photon, having a non-perturbative parton density function
(PDF).
The criterion $\mu\gg m_Q$ is certainly satisfied for $e^+e^-$ annihilation on
the $Z$-boson resonance, and for the photo-, lepto-, and hadroproduction of
$D$ and $B$ hadrons with transverse momenta $p_T\gg m_Q$.
Furthermore, the universality of the FFs is guaranteed by the factorization
theorem \cite{col}, which entitles us to transfer information on how charm and 
bottom quarks hadronize to $D$ and $B$ hadrons, respectively, in a
well-defined quantitative way from $e^+e^-$ annihilation, where the
measurements are usually most precise, to other kinds of experiments, such as
photo-, lepto-, and hadroproduction.
In Refs.~\cite{bkkc,bkkb}, the distributions in the scaled $D$- and $B$-hadron
energy $x=2E/\sqrt s$ measured at LEP1 were fitted at LO and NLO in the ZMVFNS
using, among others, the ansatz by Peterson {\it et al.} \cite{pet} for the
$c\to D$ and $b\to B$ FFs at the starting scale $\mu_0=2m_Q$.
In the $D^{*\pm}$ ($B^+/B^0$) case, the $\varepsilon$ parameter was found
to be $\varepsilon_c=0.0851$ and $0.116$ \cite{bkkc} ($\varepsilon_b=0.0126$
and $0.0198$ \cite{bkkb}) at LO and NLO, respectively.
We emphasize that the value of $\varepsilon$ carries no meaning by itself, but
it depends on the underlying theory for the description of the fragmentation
process, in particular, on the choice of the starting scale $\mu_0$, on
whether the analysis is performed in LO or NLO, and on how the final-state
collinear singularities are factorized in NLO.
An alternative to the ZMVFNS with purely non-perturbative FFs is to decompose
the FFs into a perturbative component, the so-called {\it perturbative FFs}
(PFFs) \cite{pff}, and a non-perturbative component
\cite{bkk,Cacciari:1996wr}.

In the traditional {\it massive-quark scheme} or {\it fixed-flavour-number
scheme} (FFNS), the $Q$ quark is treated in the on-mass-shell renormalization
scheme, as if it were a massive lepton in triplicate, and it only appears in
the final state, but not as an active parton inside the incoming hadron or
photon.
There are no collinear singularities associated with the outgoing $Q$-quark
lines that need to be subtracted and absorbed into FFs.
This scheme breaks down for $p_T\gg m_Q$ because of would-be collinear
singularities of the form $\alpha_s\ln\left(p_T^2/m_Q^2\right)$, which are not
resummed.
However, this scheme allows one to calculate a total cross section, which is
infeasible in the ZMVFNS.
Quantitative comparisons of the ZMVFNS and the FFNS for photoproduction in
$ep$ and $\gamma\gamma$ collisions may be found in
Refs.~\cite{Kniehl:1995em} and \cite{Cacciari:1995ej}, respectively.

A rigorous theoretical framework that retains the full finite-$m_Q$ effects
while preserving the indispensible virtues of the factorization theorem,
namely the universality and the DGLAP \cite{dglap} scaling violations of the
FFs, is provided by the {\it general-mass variable-flavour-number scheme}
(GMVFNS) \cite{col,acot}.
In a nutshell, this procedure consists in explicitly performing the $m_Q\to0$
limit of the FFNS result, comparing the outcome, term by term, with the
ZMVFNS result in the $\overline{\rm MS}$ scheme, and subtracting the
difference terms from the FFNS result.
Owing to the factorization theorem \cite{col}, the hard-scattering cross
sections thus obtained can then be convoluted with non-perturbative $D$- and
$B$-hadron FFs extracted from LEP1 data using the pure $\overline{\rm MS}$
scheme \cite{bkkc,bkkb}.
This is consistent because the finite-$m_Q$ terms omitted in
Refs.~\cite{bkkc,bkkb} are relatively small, of order $m_Q^2/m_Z^2$.
The impact of finite-$m_Q$ terms on the proton PDFs was recently assessed by
the CTEQ Collaboration \cite{Kretzer:2003it}.
In this connection, we should also mention the so-called {\it fixed-order
next-to-leading-logarithm} (FONLL) scheme \cite{fonll}, in which the ordinary
result in the FFNS and a suitably subtracted result in a ZMVFNS with PFFs are
linearly combined using a certain weight function \cite{fonll}.
The conceptual merits of the GMVFNS and the FONLL are reviewed in
Ref.~\cite{hcp}.

The GMVFNS was recently implemented for direct \cite{ks1} and single-resolved
\cite{ks2} $\gamma\gamma$ collisions as well as for $ep$ collisions with 
quasi-real photons \cite{ks3}.
In the case of $\gamma\gamma\to D^{*\pm}+X$ at LEP2, the inclusion of
finite-$m_c$ effects was found to reduce the cross section by approximately
20\% (10\%) at $p_T^D=2m_c$ ($3m_c$) \cite{ks1}, {\it i.e.}, their magnitude
is roughly $m_c^2/(p_T^D)^2$, as na\"\i vely expected.
From Refs.~\cite{ks1,ks2,ks3}, we thus infer that finite-$m_c$ effects play a
significant role only at rather small $p_T^D$ values, $p_T^D\alt3$~GeV, so
that the ZMVFNS should yield a good approximation in the kinematic range
$p_T^D>3$~GeV and $p_T^j>6$~GeV considered in a very recent ZEUS analysis
\cite{zeus}.

The article is organized as follows.
In Sec.~II, we give a short description of the method.
In Sec.~III, we present a numerical analysis of $D^{*\pm}$ plus jet associated
photoproduction, where we also compare to preliminary ZEUS data \cite{zeus}.
Section~IV contains the conclusions. 

\section{Theoretical framework}

Using the massless scheme, we can rely on the factorization theorem and write
the photoproduction cross section for $ep\to\gamma p\to D^{*\pm}+X$ as a
convolution of the partonic cross section $\hat\sigma$ with the PDFs of the
incident particles and the FFs for an outgoing parton fragmenting into a
$D^{*\pm}$ meson.
In this approach, the FFs are purely non-perturbative, universal, and subject
to DGLAP \cite{dglap} evolution.
We have
\begin{eqnarray}
d\sigma^{ep\to D^{*\pm}+X}(P_e,P_p,P_D)&=&\sum_{i,j,k}\int dx_e d x_p dz\,
F_{i/e}(x_e,M_e)F_{j/p}(x_p,M_p)\,D_{D/k}(z,M_F)\nonumber\\
&&{}\times d\hat\sigma^{ij\to kX}(x_eP_e,x_pP_p,P_D/z,\mu,M_e,M_p,M_F),
\label{dsigma}
\end{eqnarray}
where $M_e$ and $M_p$ are the initial-state factorization scales, $M_F$ is the 
final-state factorization scale, and $\mu$ the renormalization scale. 
If a jet in addition to the $D^{*\pm}$ meson is detected, a measurement
function defining the jet has to be included in Eq.~(\ref{dsigma}).
The sum $\sum_{i,j,k}$ runs over all partons, including quarks, gluons as well
as photons, which can contribute if the energy of the subprocess is above
their mass thresholds. 
Therefore, the charm quark also contributes as an incoming parton originating
from the proton or the resolved photon. 

\begin{figure}[htb]
\begin{picture}(80,40)(30,-5)
\put(0,0){\mbox{\epsfig{file=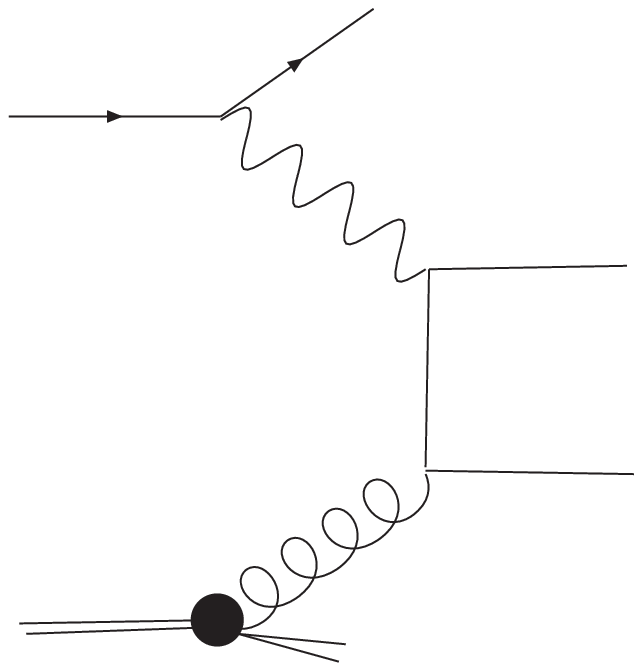,height=3cm}}}
\put(0,12){(a)}
\put(50,0){\mbox{\epsfig{file=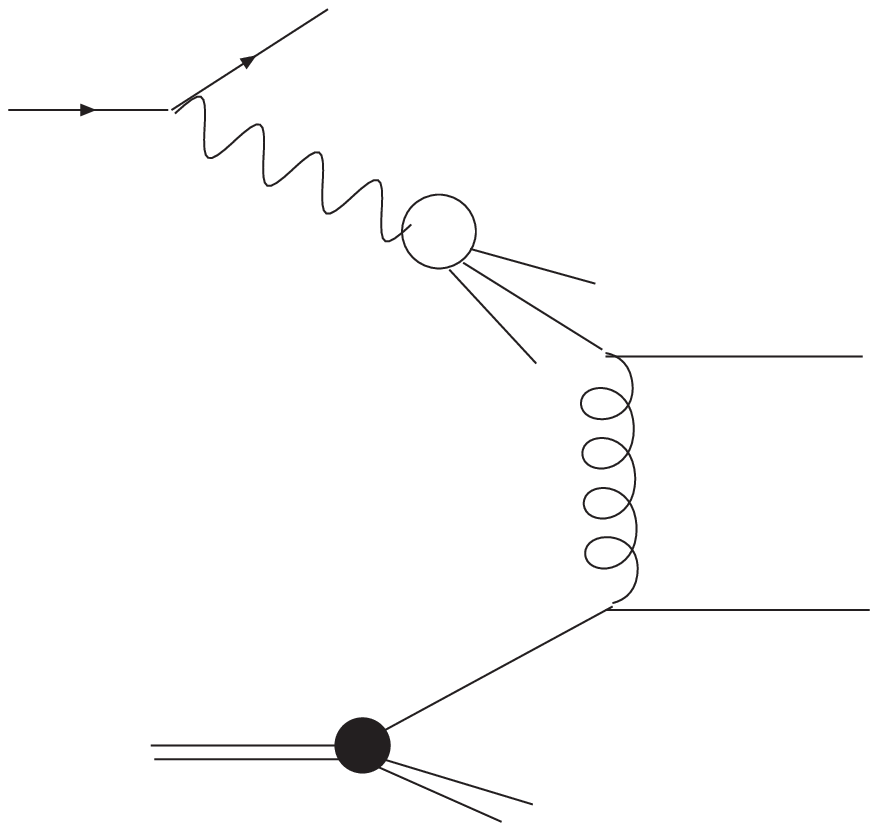,height=3cm}}}
\put(50,12){(b)}
\put(100,0){\mbox{\epsfig{file=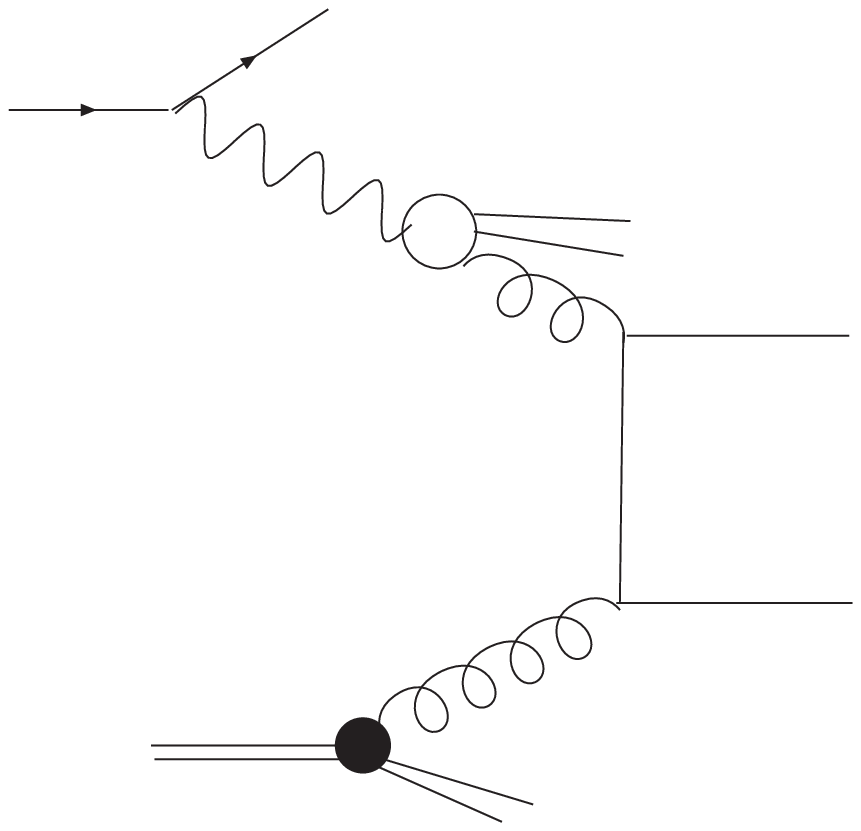,height=3cm}}}
\put(100,12){(c)}
\end{picture}
\caption{Examples of LO diagrams involving (a) direct photons or (b) the quark
or (c) gluon components of resolved photons.}
\label{fig1}
\end{figure}

At leading order, the subprocesses contributing to the partonic reaction
$ij\to k X$ can be divided into two categories, corresponding to a direct or a
resolved photon in the initial state. 
The direct photon corresponds to $i=\gamma$ in Eq.~(\ref{dsigma}), with
$F_{\gamma/e}$ approximated by the Weizs\"acker-Williams formula for the
spectrum of the quasi-real photons,
\begin{equation}
F_{\gamma/e}(y) = \frac{\alpha_{em}}{2\pi}\left[\frac{1+(1-y)^2}{y}
\ln{\frac{Q^2_{\rm max}(1-y)}{m_e^2y^2}}-\frac{2(1-y)}{y}\right],
\label{ww}
\end{equation}
where $\alpha_{em}$ is Sommerfeld's fine-structure constant and
$Q^2_{\rm max}$ is the maximum virtuality of the photon.
On the other hand, the resolved photon participates in the hard interaction 
via its quark and gluon content.
In this case, $F_{i/e}(x_e,M_e)$ is given by a convolution of the 
Weizs\"acker-Williams spectrum with the photon PDFs as
\begin{equation}
F_{i/e}(x_e,M_e)=\int_0^1 dy dx_\gamma\,F_{\gamma/e}(y)
F_{i/\gamma}(x_\gamma,M_e)\delta(yx_\gamma-x_e).
\end{equation}
Typical examples of LO diagrams are depicted in Fig.~\ref{fig1}.

\begin{figure}[htb]
\begin{picture}(80,40)(30,-5)
\put(0,0){\mbox{\epsfig{file=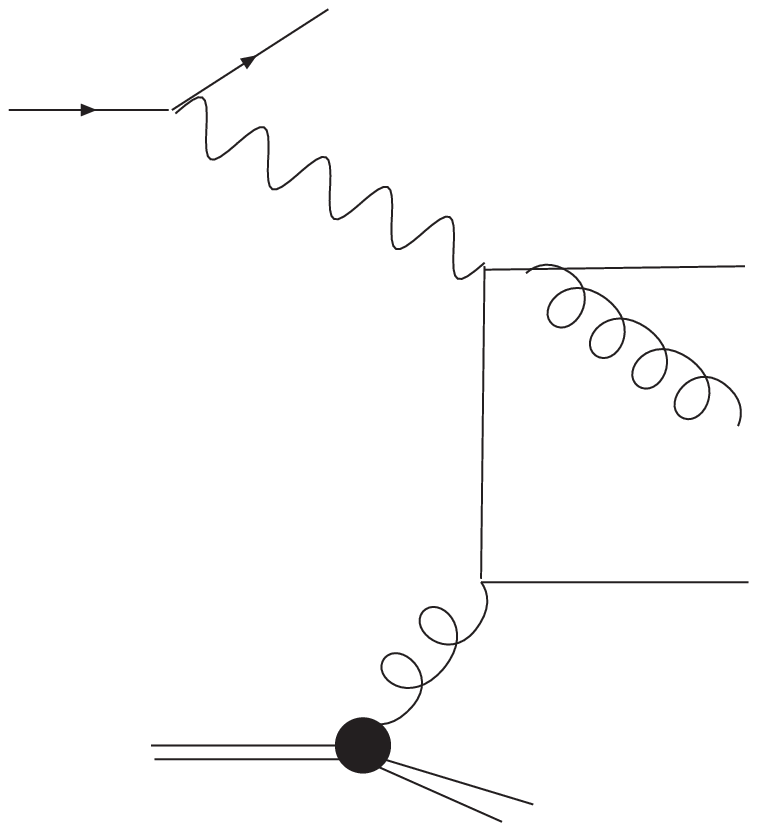,height=3cm}}}
\put(0,12){(a)}
\put(50,0){\mbox{\epsfig{file=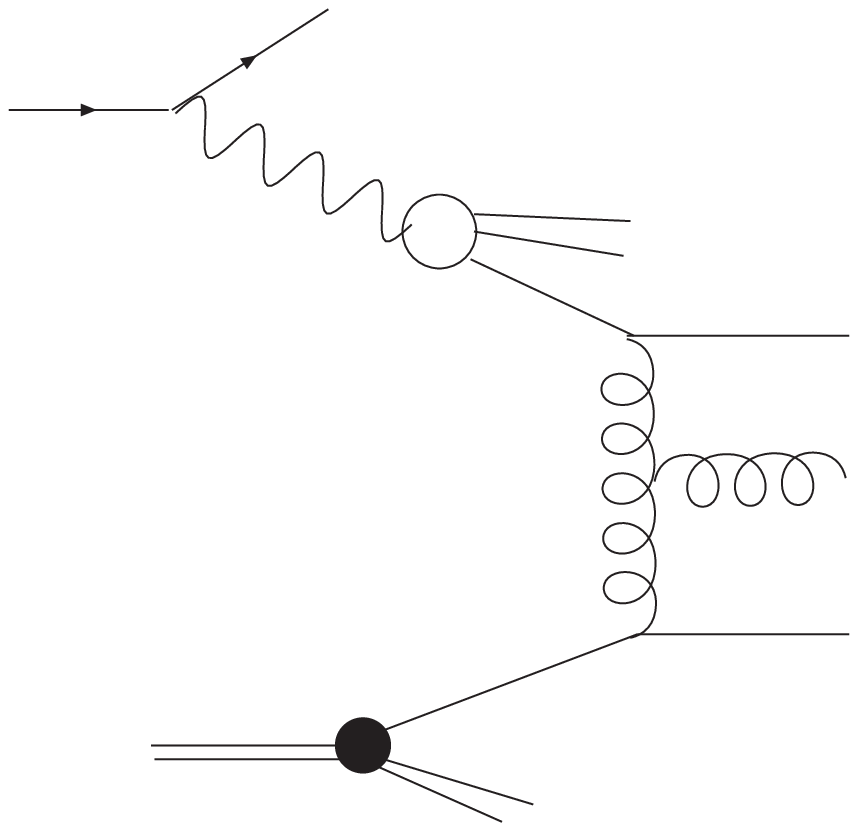,height=3cm}}}
\put(50,12){(b)}
\put(100,0){\mbox{\epsfig{file=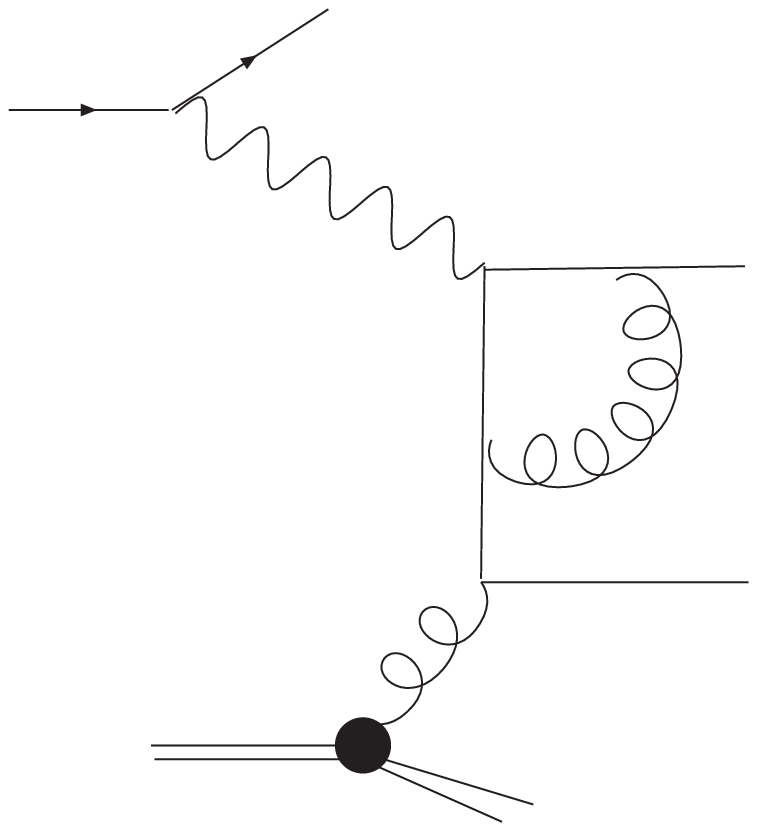,height=3cm}}}
\put(100,12){(c)}
\end{picture}
\caption{Examples of NLO diagrams contributing to the real corrections in
(a) direct and (b) resolved photoproduction, and (c) to the virtual
corrections in direct photoproduction.}
\label{fig2}
\end{figure}

At NLO, an additional real or virtual parton can be radiated in the hard
interaction.
Representative examples of NLO diagrams are shown in Fig.~\ref{fig2}. 
As the partons are massless, this leads to soft and collinear singularities.
The soft singularities cancel between the real and the virtual corrections,
while the collinear ones are absorbed into the PDFs or FFs. 
For example, the collinear singularities appearing at NLO if the incident
photon splits into a collinear $q\bar q$ pair are absorbed into the PDF
$F_{q/\gamma}(x_\gamma,M_e)$ at the factorization scale $M_e$.
Thus the direct- and resolved-photon contributions separately exhibit strong
dependences on $M_e$, which cancel out only in their sum, up to terms which
are formally beyond NLO.
Therefore, it has to be stressed that, at NLO, only the sum of the two 
contributions carries a physical meaning.  
Technically, the infrared divergences have been isolated using a combination
of the phase-space-slicing \cite{slicing} and the subtraction
\cite{subtraction} methods.
A more detailed description of the calculation can be found in
Refs.~\cite{Fontannaz,Fontannaz:2002nu} and will not be repeated here. 
The matrix elements have been calculated in Refs.~\cite{Aurenche:1984hc} and
are implemented in the computer program {\tt EPHOX} \cite{ephox}, which can
serve to calculate the photoproduction of a large-$p_T$ hadron or photon
together with an optional jet.
The program is constructed as a partonic event generator, so that the total
cross section as well as differential distributions can be easily obtained. 

\section{Numerical results}

We are now in a position to present our numerical analysis.
We start by specifying our inputs and the kinematic situation. 
We work in the ZMVFNS with $n_f=5$ massless quark flavors.
For the PDFs of the proton, we take the MRST03 \cite{mrst3} set by Martin,
Roberts, Stirling, and Thorne.
For the PDFs of the photon, our default set is AFG04 \cite{afg04} by Aurenche,
Fontannaz, and Guillet, which is an updated version of the original AFG
\cite{Aurenche:1994in} parameterization.
In contrast to the AFG \cite{Aurenche:1994in} set, the AFG04 \cite{afg04} set
also contains a bottom-quark PDF.
The AFG04 \cite{afg04} PDFs are slightly higher at small values of $x$ and
lower at large values of $x$ than the AFG \cite{Aurenche:1994in} PDFs, but the
numerical difference is very small. 
In order to assess the potential of the ZEUS data \cite{zeus} to constrain the
photon PDFs, we also employ set GRV~HO \cite{grv} by Gl\"uck, Reya, and Vogt,
which we transform from the DIS$_\gamma$ scheme to the $\overline{\rm MS}$
scheme. 
For the $D^{*\pm}$ FFs, we use the parameterization of Ref.~\cite{bkkc}, where
separate NLO fits to ALEPH and OPAL data are performed.
As our default, we chose the set obtained from the fit to the OPAL data, as it
has a lower $\chi^2$ value.
However, the differences in the considered cross sections resulting from
exchanging the two FF sets are negligible.
For $\alpha_s^{(n_f)}(\mu)$, we use an exact solution of the two-loop
renormalization group equation, where the asymptotic scale parameter
$\Lambda^{(5)}$ for $n_f=5$ is calculated from $\Lambda^{(4)}$ by requiring
continuity of $\alpha_s$ at the bottom threshold $\mu=2m_b$, and
$\Lambda^{(4)}$ is set to 278~MeV to be consistent with the MRST03
\cite{mrst3} proton PDFs.

\begin{figure}[htb]
\begin{center}
\mbox{\epsfig{file=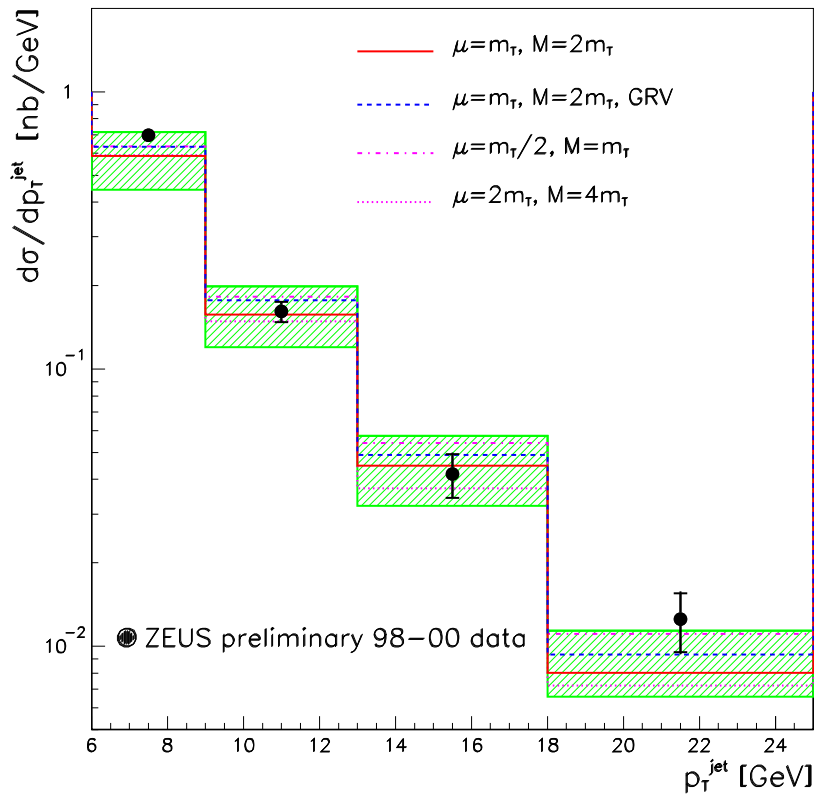,height=12.cm}}
\end{center}
\caption{The differential cross section $d\sigma/dp_T^j$ of
$ep\to D^{*\pm}j+X$ in photoproduction for $-1.5<\eta^j<2.4$, $p_T^D>3$~GeV,
and $|\eta^D|<1.5$ measured by ZEUS \cite{zeus} is compared with our NLO
predictions (solid red histogram) including conservative errors (hatched green
bands).
For comparison, the errors due to diagonal scale variation (dotted-dashed 
and dotted pink histograms) and the predictions evaluated for central scale
choice with the GRV~HO photon PDFs (dashed blue histogram) are also shown.}
\label{ptscal}
\end{figure}

\begin{figure}[htb]
\begin{center}
\mbox{\epsfig{file=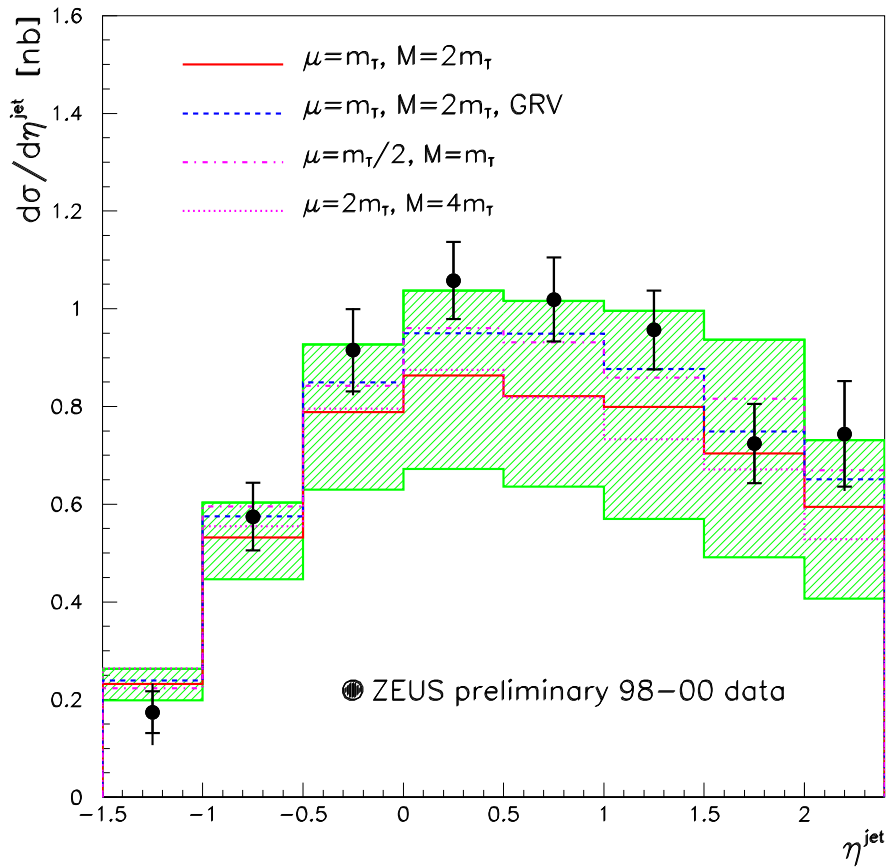,height=13cm}}
\end{center}
\caption{Same as in Fig.~\ref{ptscal}, but for $d\sigma/d\eta^j$ with
$p_T^j>6$~GeV.}
\label{rapscal}
\end{figure}

\begin{figure}[htb]
\begin{picture}(200,70)(15,5)
\put(0,0){\mbox{\epsfig{file=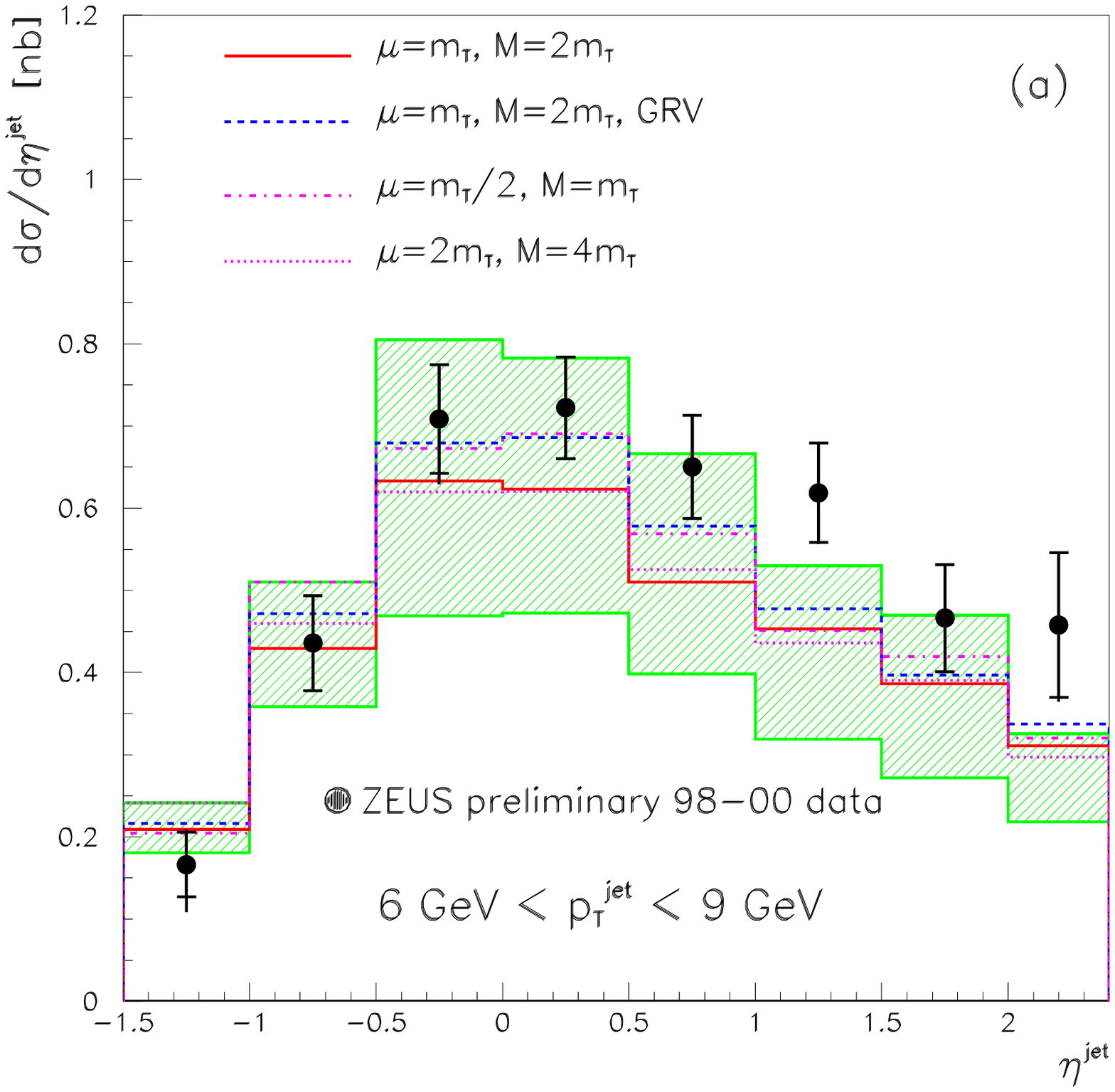,height=9.8cm}}}
\put(92,0){\mbox{\epsfig{file=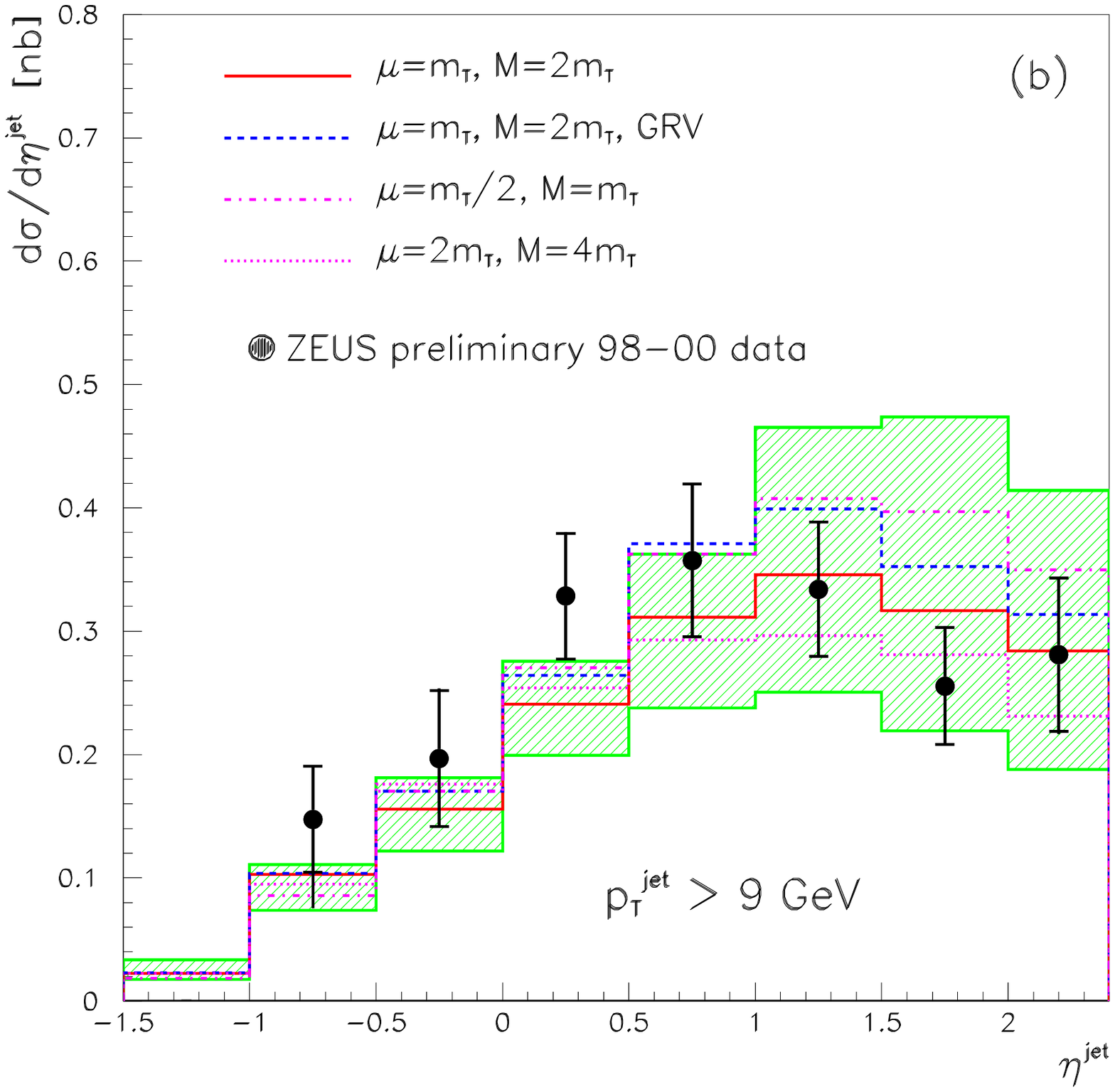,height=9.8cm}}}
\end{picture}
\caption{Same as in Fig.~\ref{rapscal}, but for (a) $6<p_T^j<9$~GeV and (b)
$p_T^j>9$~GeV.}
\label{raplow}
\end{figure}

\begin{figure}[htb]
\begin{picture}(200,200)(15,-100)
\put(0,0){\mbox{\epsfig{file=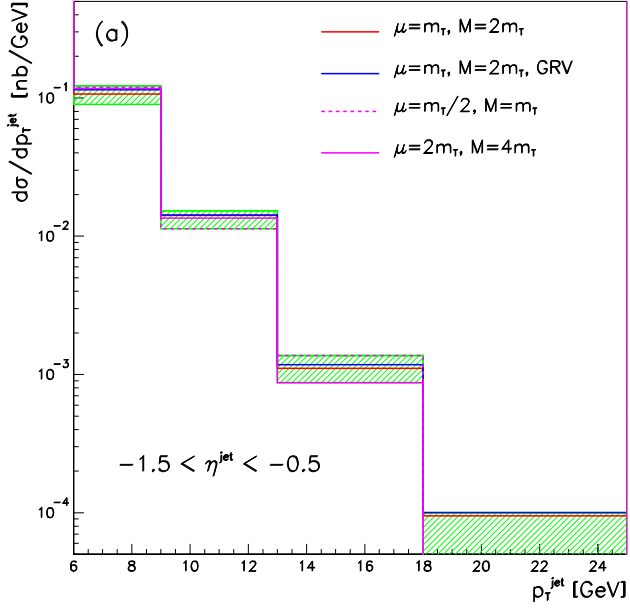,height=9.2cm}}}
\put(100,0){\mbox{\epsfig{file=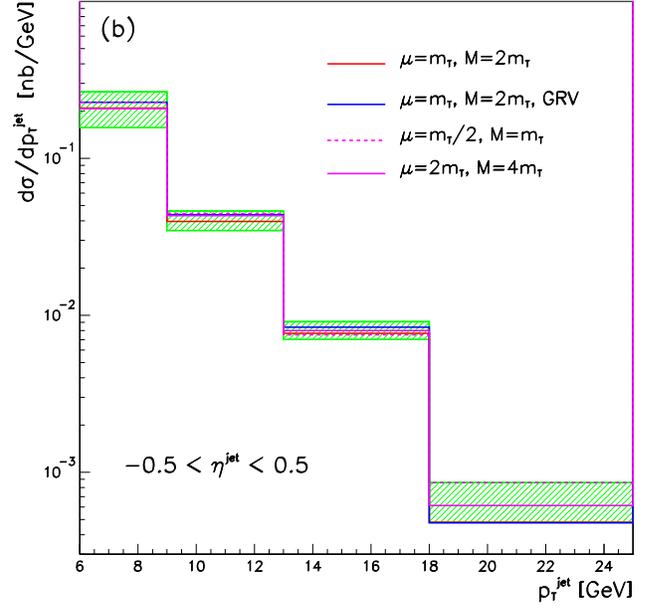,height=9.2cm}}}
\put(0,-100){\mbox{\epsfig{file=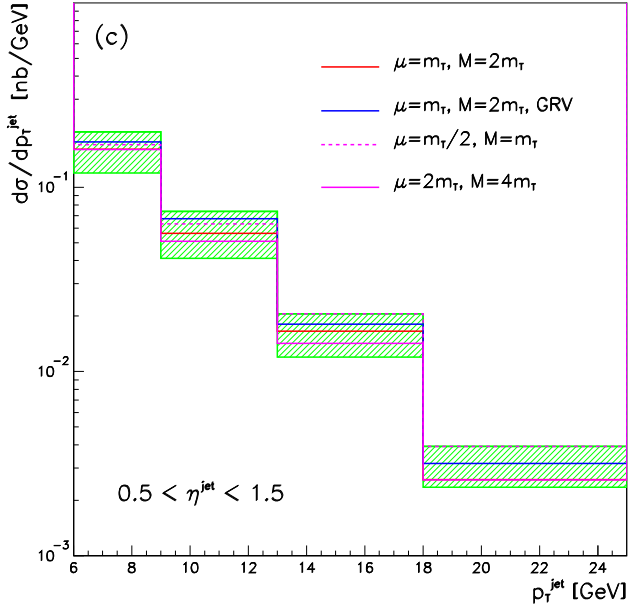,height=9.2cm}}}
\put(100,-100){\mbox{\epsfig{file=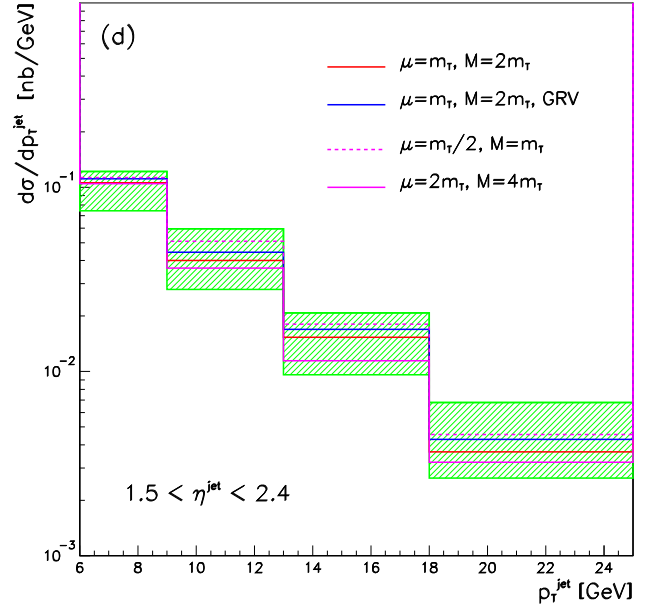,height=9.2cm}}}
\end{picture}
\caption{Same as in Fig.~\ref{ptscal}, but for (a) $-1.5<\eta^j<-0.5$, (b)
$-0.5<\eta^j<0.5$, (c) $0.5<\eta^j<1.5$, and (d) $1.5<\eta^j<2.4$ and without
experimental data.}
\label{ptrapcuts}
\end{figure}

Our default scale choice is $\mu=m_T$ and $M=2m_T$,
where $m_T=\sqrt{m_c^2+(p_T^D)^2}$ is the transverse mass of the $D^{*\pm}$
meson and we set $m_c=1.5$~GeV.
As usual, we identify the three factorization scales $M_e$, $M_p$, and $M_F$
in Eq.~(\ref{dsigma}) and denote their common value by $M$. 
In order to estimate the theoretical uncertainty in a conservative way, we
vary $\mu$ and $M$ independently about their default values, as $\mu=c_1m_T$
and $M=2c_2m_T$ with $1/2\leq c_1,c_2\leq2$.
The maximum variation of the cross section thus obtained is shown as hatched
green bands in Figs.~\ref{ptscal}--\ref{ptrapcuts}.
For comparison, we also consider {\it diagonal} scale variations, where
$c_1=c_2$, in these figures.
However, we caution the reader that the errors resulting from diagonal scale 
variations are too optimistic because of cancellations due to the fact that
the dependences of the cross section on $\mu$ and $M$ act in opposite
directions. 

We chose the kinematics in such a way that a direct comparison to preliminary
ZEUS data on $D^{*\pm}$ plus jet associated photoproduction \cite{zeus} is
possible.
These data have been produced with a proton energy of 920~GeV and an electron
energy of 27.5~GeV in the laboratory frame, which corresponds to a $ep$
center-of-mass (c.m.) energy of $\sqrt s=318$~GeV. 
The maximum photon virtuality is $Q^2_{\rm max}=1$~GeV$^2$, and the 
photon-proton c.m.\ energy $W$ lies in the range 130~GeV${}<W<280$~GeV.
All rapidities $\eta$ refer to the laboratory frame, with the HERA convention
that the proton is moving towards positive rapidity.
The jets are defined using the $k_T$-algorithm \cite{ktalgo} with a jet radius
of $R=1$.
We employ the ZEUS convention \cite{zeus} that an event is counted twice if
two jets in addition to the $D^{*\pm}$ meson are measured which both satisfy
the cuts in transverse momentum and rapidity.
However, we find that this case only occurs for about 0.3\% of the generated
events in the kinematic range considered.
Unless stated otherwise, we adopt the acceptance cuts $p_T^D>3$~GeV,
$|\eta^D|<1.5$, $p_T^j>6$~GeV, and $-1.5<\eta^j<2.4$ from ZEUS \cite{zeus}.
We present histograms with the same binning as in Ref.~\cite{zeus}.
As in Ref.~\cite{zeus}, we consider the sum of the cross sections for $D^{*+}$
and $D^{*-}$ mesons.

We now present and discuss our figures.
In Figs.~\ref{ptscal}--\ref{raplow}, we confront the differential cross
section of $ep\to D^{*\pm}j+X$ in photoproduction as measured by ZEUS 
\cite{zeus} with our NLO predictions.
Specifically, Fig.~\ref{ptscal} refers to $d\sigma/dp_T^j$ integrated over the
full $\eta^j$ range ($-1.5<\eta^j<2.4$) and Figs.~\ref{rapscal},
\ref{raplow}(a) and (b) to $d\sigma/d\eta^j$ integrated over the full $p_T^j$
range ($p_T^j>6$~GeV) and the subintervals $6<p_T^j<9$~GeV and $p_T^j>9$~GeV,
respectively.
In addition, we present, in Figs.~\ref{ptrapcuts}(a)--(d), $d\sigma/dp_T^j$
integrated over the subintervals $-1.5<\eta^j<-0.5$, $-0.5<\eta^j<0.5$,
$0.5<\eta^j<1.5$, and $1.5<\eta^j<2.4$, for which experimental data are not
yet available.
In all cases, the $D^{*\pm}$ meson is kinematically confined by the conditions
$p_T^D>3$~GeV and $|\eta^D|<1.5$.
The vertical bars on the ZEUS data \cite{zeus} give the full errors, the
purely statistical errors are indicated by the horizontal ticks on them.
Our default predictions are represented by the solid red histograms and their
errors due to independent scale variation by the hatched green bands.
The errors due to diagonal scale variation are indicated by the dot-dashed and
dotted pink histograms.
The predictions evaluated for central scale choice with the GRV~HO photon PDFs
are given by the dashed blue histograms.
We observe that our NLO predictions agree with the ZEUS data \cite{zeus}
within errors, except for two forward $\eta^j$ bins in the lower $p_T^j$
range, where the ZEUS data \cite{zeus} slightly overshoot our NLO predictions.
In the cases when the difference between the evaluations with the AFG04
\cite{afg04} and GRV~HO \cite{grv} photon PDFs is comparable to the
experimental error, the GRV~HO \cite{grv} set tends to yield a better
description of the ZEUS data \cite{zeus}, except at large values of $p_T^j$
and $\eta^j$.

\begin{figure}[htb]
\begin{picture}(200,80)(20,0)
\put(50,0){\mbox{\epsfig{file=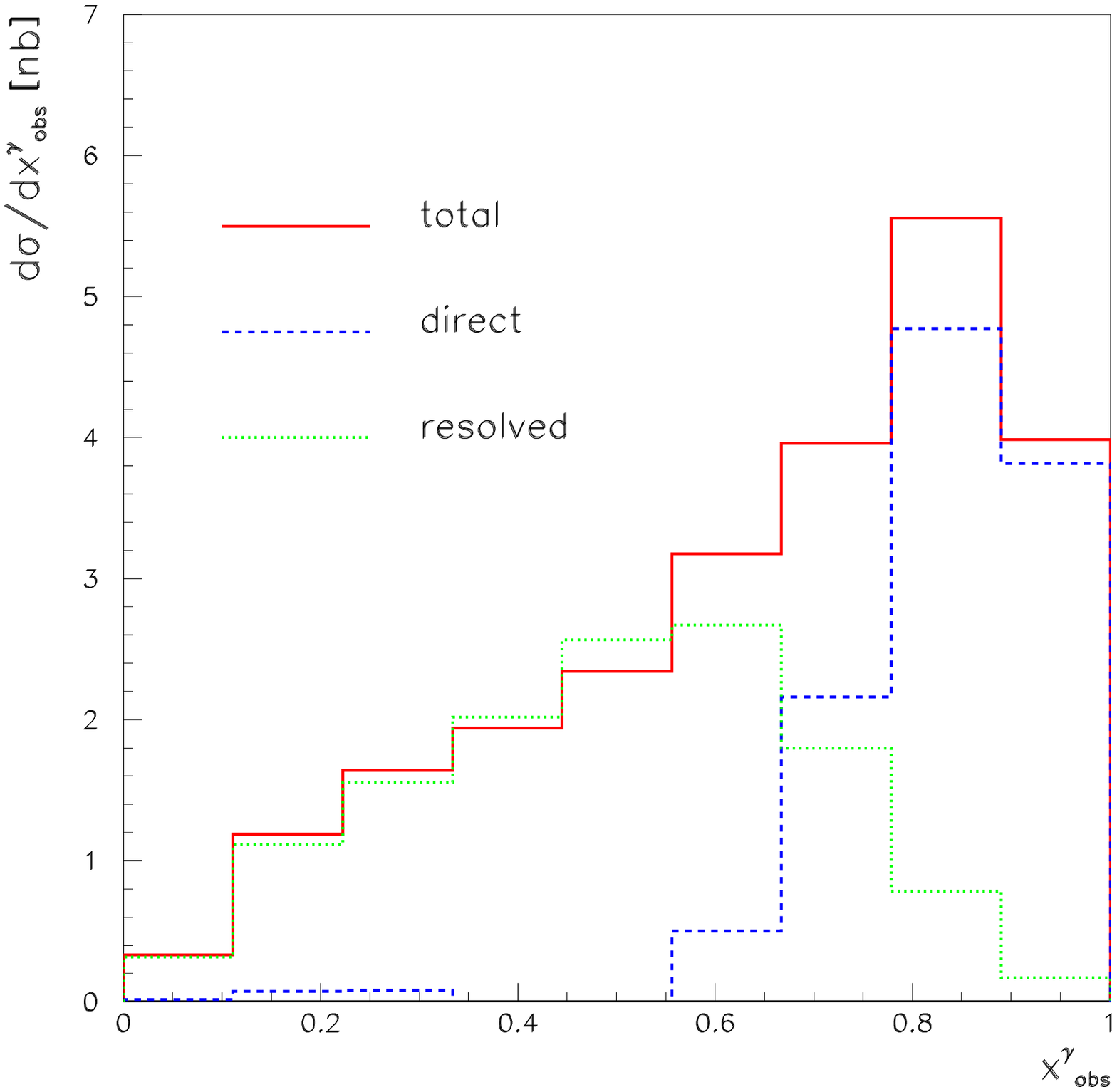,height=8.5cm}}}
\end{picture}
\caption{Our central NLO prediction of the differential cross section
$d\sigma/dx_{\rm obs}^\gamma$ of $ep\to D^{*\pm}j+X$ in photoproduction for
$p_T^D>3$~GeV, $|\eta^D|<1.5$, $p_T^j>6$~GeV, and $-1.5<\eta^j<2.4$ (solid red
histogram) and its direct- (dashed blue histogram) and  resolved-photon
(dotted green histogram) components.}
\label{figxobs}
\end{figure}

It is also instructive to look at the distribution 
$d\sigma/dx_{\rm obs}^\gamma$, where the kinematic observable
$x_{\rm obs}^\gamma$ is defined as 
\begin{equation}
x_{\rm obs}^\gamma=\frac{p_T^D\exp(-\eta^D)+p_T^j\exp(-\eta^j)}
{2E^\gamma}.
\label{xobs}
\end{equation} 
As the contribution from the direct-photon subprocesses peaks at
$x_{\rm obs}^\gamma\approx1$, whereas resolved photons mainly contribute for
$x_{\rm obs}^\gamma<1$, a cut on $x_{\rm obs}^\gamma$ can serve to obtain
samples enriched in direct- or resolved-photon processes. 
As already mentioned in Sec.II, the true direct- and resolved-photon
contributions are related at NLO through factorization and, taken separately,
exhibit strong $M_e$ dependences.
Only their sum represents a physical observable that can be compared to 
experimental data.
Notice that the individual parts can even be negative.
In Fig.~\ref{figxobs}, our central NLO prediction for the differential cross
section $d\sigma/dx_{\rm obs}^\gamma$ in the kinematic range $p_T^D>3$~GeV,
$|\eta^D|<1.5$, $p_T^j>6$~GeV, and $-1.5<\eta^j<2.4$ (solid red histogram) is
decomposed into its direct- (dashed blue histogram) and resolved-photon
(dotted green histogram) components.
We read off from Fig.~\ref{figxobs} that, with our default scale choice,
direct photoproduction dominates for $x_{\rm obs}^\gamma\agt0.7$.

\begin{figure}[htb]
\begin{picture}(200,80)(5,0)
\put(0,0){\mbox{\epsfig{file=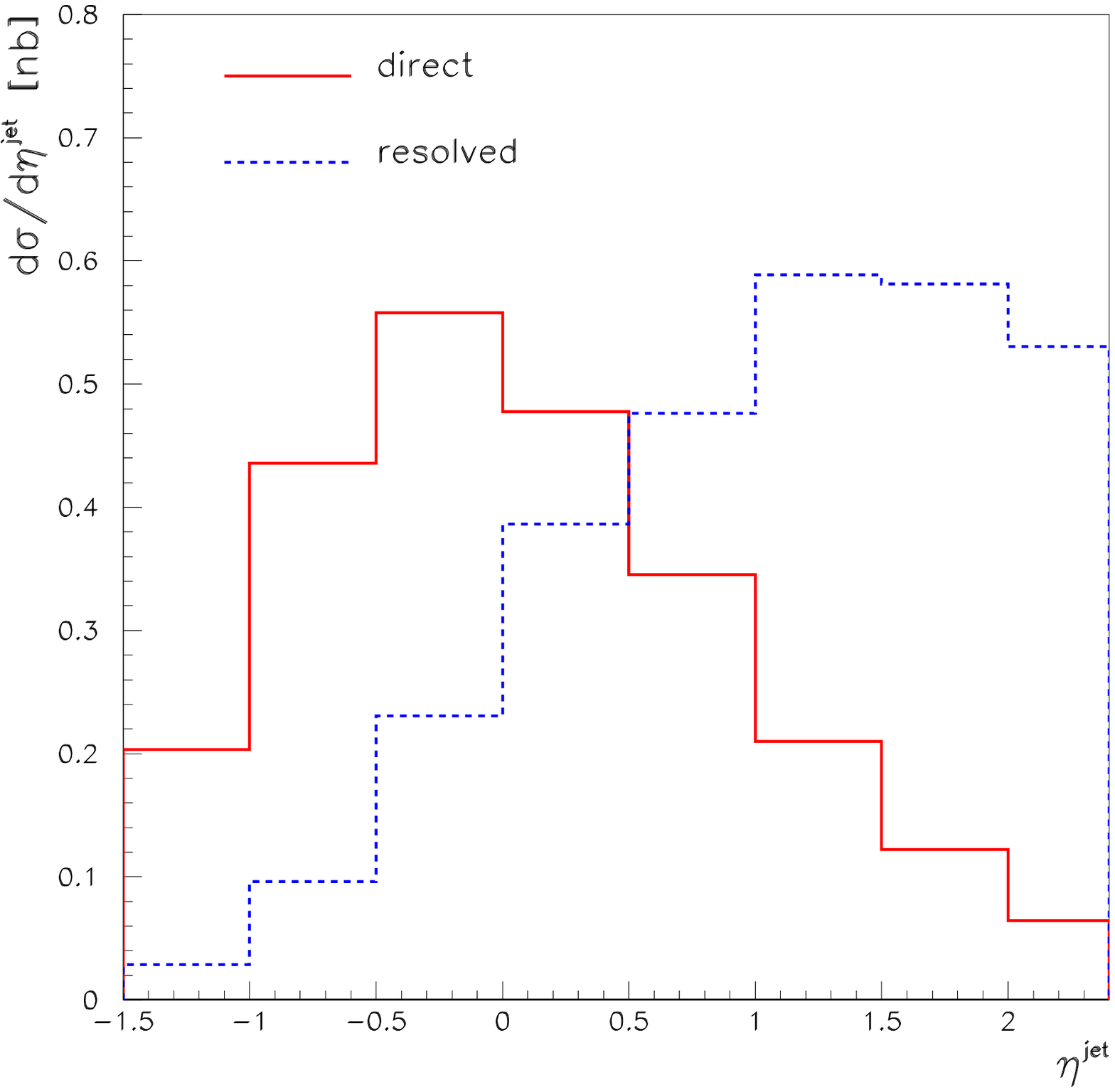,height=8.5cm}}}
\put(90,0){\mbox{\epsfig{file=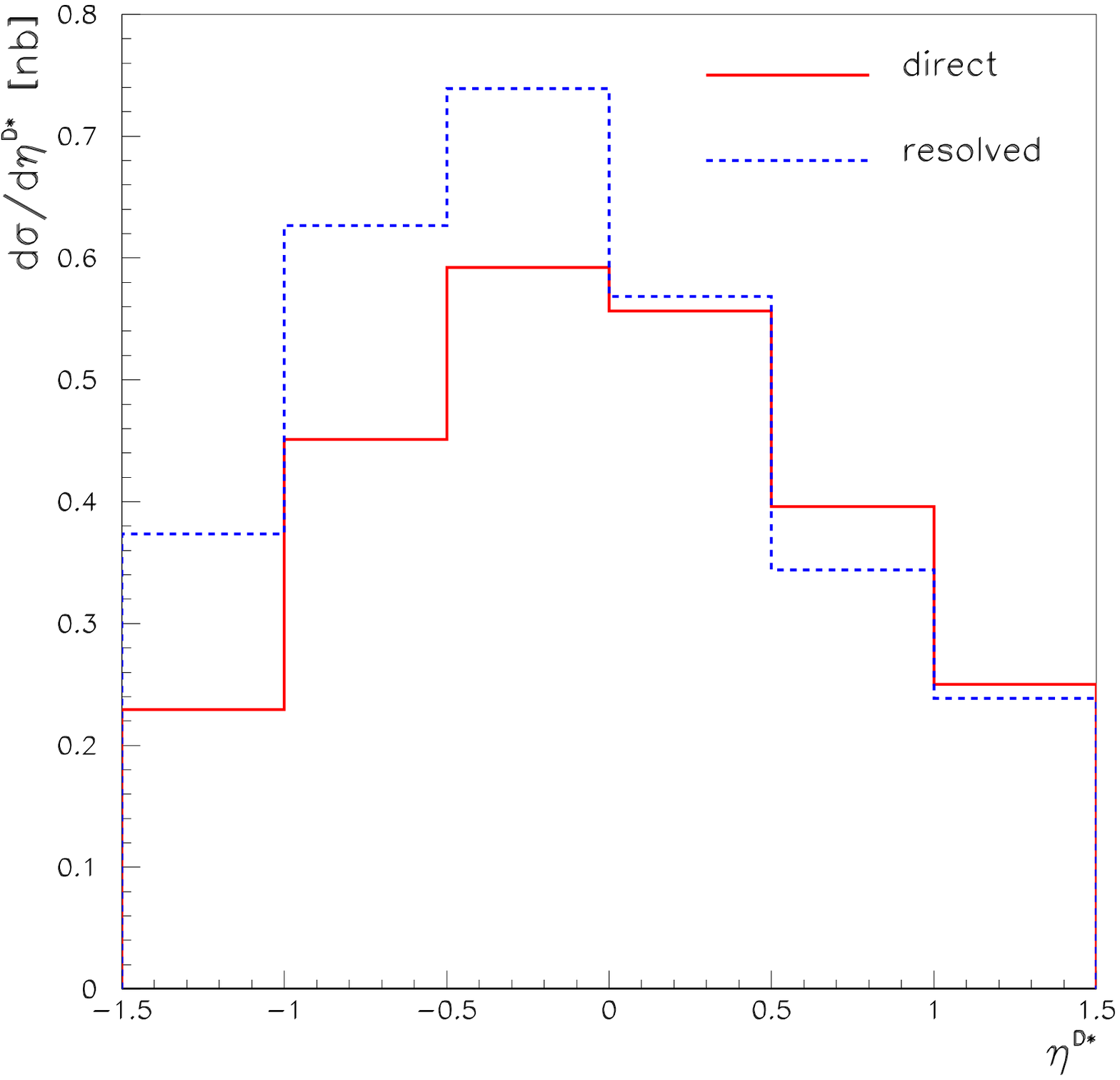,height=8.5cm}}}
\end{picture}
\caption{Direct- (solid red histograms) and  resolved-photon (dashed blue
histograms) components of our central NLO predictions of the differential
cross sections (a) $d\sigma/d\eta^j$ for $p_T^D>3$~GeV, $|\eta^D|<1.5$, and
$p_T^j>6$~GeV and (b) $d\sigma/d\eta^D$ for $p_T^D>3$~GeV, $p_T^j>6$~GeV, and
$-1.5<\eta^j<2.4$ of $ep\to D^{*\pm}j+X$ in photoproduction.}
\label{figresodir}
\end{figure}

In the case of single-hadron inclusive photoproduction at HERA, the direct-
and resolved-photon contributions are known to be accumulated in the backward
and forward directions, respectively \cite{bor}.
It is interesting to find out if, in $D^{*\pm}$ plus jet associated
photoproduction, the rapidities $\eta^D$ and $\eta^j$ lend themselves as
discriminators between direct and resolved photoproduction as well.
To this end, we split our central NLO predictions for the differential cross
sections $d\sigma/d\eta^j$ and $d\sigma/d\eta^D$ into their direct- and
resolved-photon components and show the results in Figs.~\ref{figresodir}(a)
and (b) as the solid red and dashed blue histograms, respectively.
The kinematic range considered for $d\sigma/d\eta^j$ is $p_T^D>3$~GeV,
$|\eta^D|<1.5$, and $p_T^j>6$~GeV, while for $d\sigma/d\eta^D$ it is
$p_T^D>3$~GeV, $p_T^j>6$~GeV, and $-1.5<\eta^j<2.4$.
Notice that the superposition of the two histograms in
Fig.~\ref{figresodir}(a) yields the solid histogram in Fig.~\ref{rapscal}.
We learn from Figs.~\ref{figresodir}(a) and (b) that, in the kinematic range 
considered, the discriminating power of the rapidity distribution with respect
to direct and resolved photons, which is familiar from single-hadron inclusive
photoproduction at HERA, carries over to $d\sigma/d\eta^j$, but not to
$d\sigma/d\eta^D$.

\begin{figure}[htb]
\begin{picture}(200,110)(20,0)
\put(50,0){\mbox{\epsfig{file=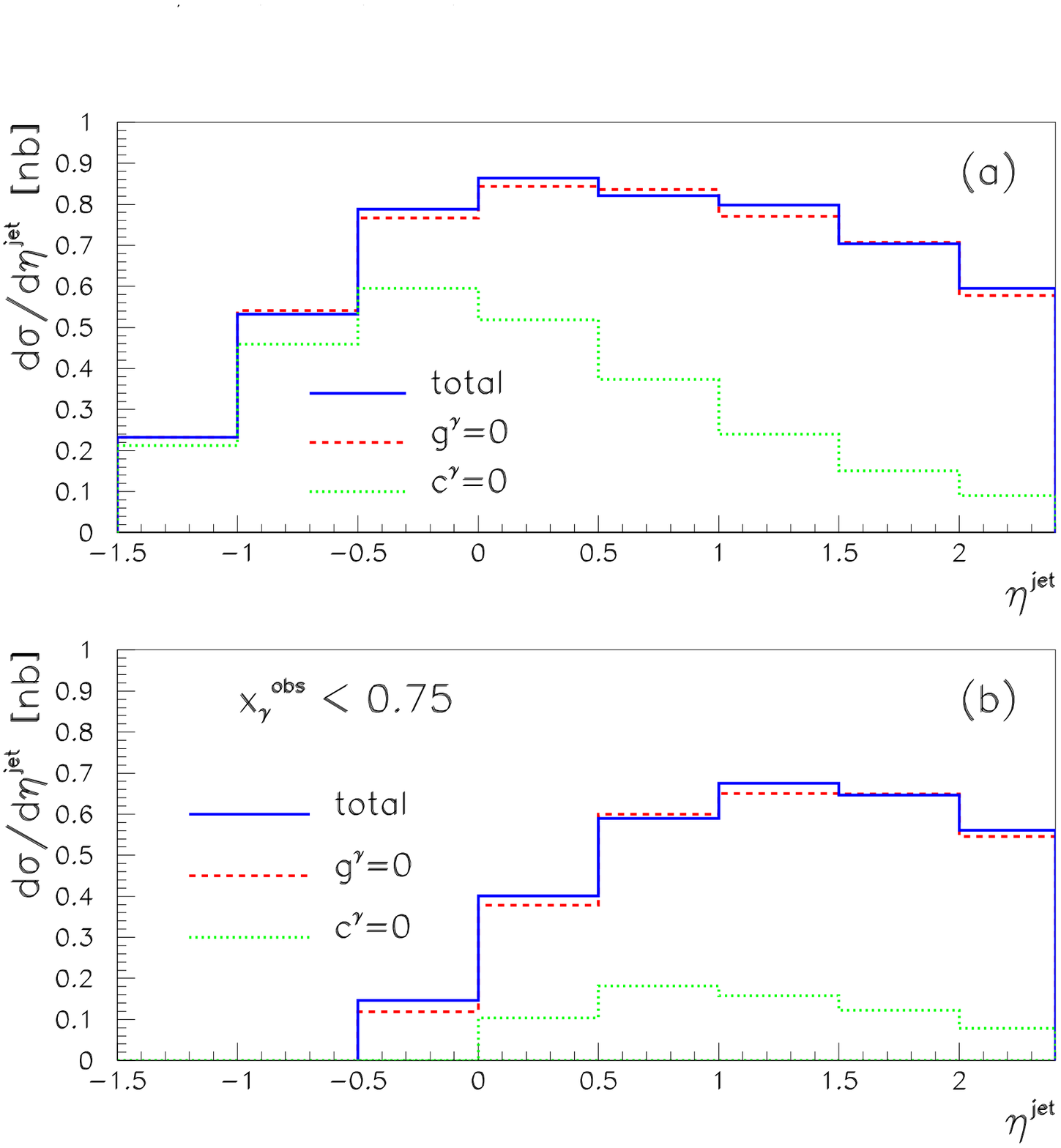,height=12cm}}}
\end{picture}
\caption{Our central NLO prediction of the differential cross section
$d\sigma/d\eta^j$ of $ep\to D^{*\pm}j+X$ in photoproduction for
$p_T^D>3$~GeV, $|\eta^D|<1.5$, and $p_T^j>6$~GeV (a) without and (b) with the
acceptance cut $x_{\rm obs}^\gamma<0.75$ (solid blue histograms) and the
evaluations with $F_{g/\gamma}=0$ (dashed red histograms) or $F_{c/\gamma}=0$
(dotted green histograms).}
\label{figlug0}
\end{figure}

We now explore the sensitivity of our NLO predictions of $ep\to D^{*\pm}j+X$
in photoproduction to the gluon and charm-quark PDFs of the photon, so as to
assess the potential of the HERA experiments to constrain these PDFs, which
are presently less well known than those of the up, down, and strange quarks.
In order to enhance this sensitivity, it is useful to suppress the
direct-photon contribution.
From the discussion of Figs.~\ref{figxobs} and \ref{figresodir}, we know that
this can be achieved by focusing on low values values of $x_{\rm obs}^\gamma$
and/or large values of $\eta^j$, typically $x_{\rm obs}^\gamma\alt0.75$ and
$\eta^j\alt0.5$.
Therefore, we reconsider in Fig.~\ref{figlug0} the differential cross section
$d\sigma/d\eta^j$ for $p_T^D>3$~GeV, $|\eta^D|<1.5$, and $p_T^j>6$~GeV (a)
without and (b) with the acceptance cut $x_{\rm obs}^\gamma<0.75$ (solid blue
histograms) and turn off the gluon (dashed red histograms) and charm-quark
(dotted green histograms) PDFs, one at a time.
Notice that the solid histograms in Figs.~\ref{rapscal} and \ref{figlug0}(a)
are identical.
We caution the reader that, strictly speaking, it is inconsistent to put to
zero a photon PDF by hand because, in the determination of the PDF set, it
participated in the DGLAP \cite{dglap} evolution and thus influenced the other
PDFs.
Furthermore, its $M_e$ dependence is correlated with a similar $M_e$
dependence in the direct-photon contribution through the factorization
procedure.
Bearing these caveats in mind, it is nevertheless instructive to do so.
We conclude from Fig.~\ref{figlug0}(a) and (b) that the contribution from the
gluon inside the resolved photon is too small to be useful, while that from
the charm quark is very significant.
Quantitatively comparing Figs.~\ref{rapscal}, \ref{figresodir}(a), and
\ref{figlug0}(a) after integration over $\eta^j$, we find that the 
contribution due to the charm component in the photon makes up 92\% of the
resolved-photon contribution and 50\% of the total cross section.
After imposing the condition $x_{\rm obs}^\gamma<0.75$, the latter fraction is 
increased to as much as 81\%!
In fact, the ZEUS data \cite{zeus} in Fig.~\ref{rapscal} overshoot the dotted
green histogram in \ref{figlug0}(a) by several experimental standard
deviations in the upper $\eta^j$ bins, which suggest that, in the framework of
the ZMVFNS, the existence of intrinsic charm in the resolved photon is
experimentally established.

\begin{figure}[htb]
\begin{center}
\mbox{\epsfig{file=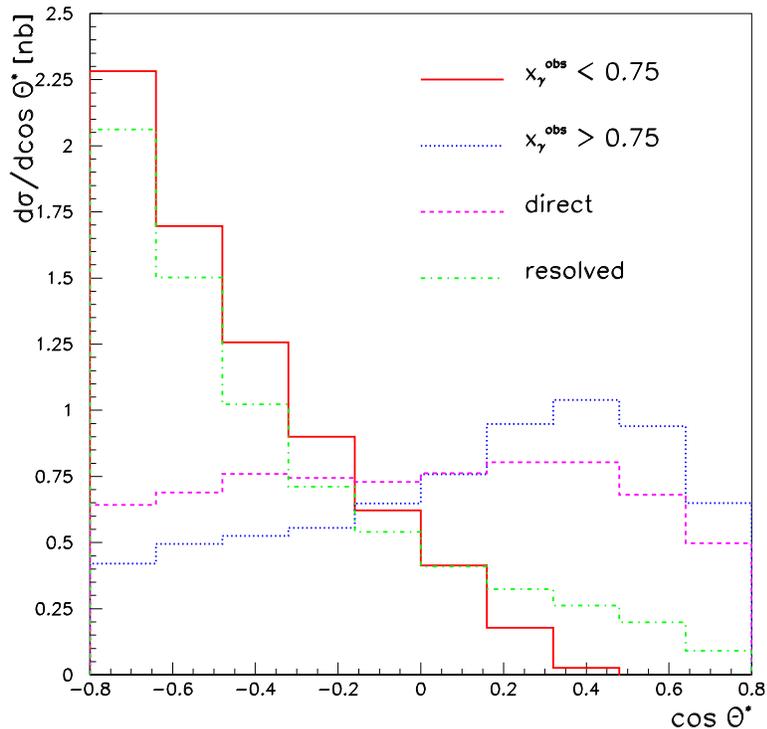,height=11cm}}
\end{center}
\caption{Direct- (dashed pink histogram) and  resolved-photon (dot-dashed
green histogram) contributions as well as contributions with
$x_{\rm obs}^\gamma>0.75$ (dotted blue histogram) and
$x_{\rm obs}^\gamma<0.75$ (solid red histogram) to our central NLO prediction
of the differential cross section $d\sigma/d\cos^*$ of $ep\to D^{*\pm}j+X$ in
photoproduction for $p_T^D>3$~GeV, $|\eta^D|<1.5$, $p_T^j>6$~GeV, and
$-1.5<\eta^j<2.4$.}
\label{cos}
\end{figure}

Another interesting observable is $\cos\theta^*$, defined as
\begin{equation}
\cos\theta^*=\tanh\frac{\eta^D-\eta^j}{2}.
\end{equation}
As the angular dependence of subprocesses involving a gluon propagator in the
$t$ channel is approximately proportional to $(1-|\cos\theta^*|)^{-2}$,
whereas it is proportional to $(1-|\cos\theta^*|)^{-1}$ in the case of a quark
propagator, one can learn about the size of the contribution from diagrams of
the type shown in Fig.~\ref{fig1}(b) by studying the differential cross
section $d\sigma/d\cos\theta^*$.
A recent ZEUS analysis on dijet angular distributions in the photoproduction
of open charm \cite{Chekanov:2003bu} has shown that the measured cross section
from {\it resolved-enriched} events, with $x_{\rm obs}^\gamma<0.75$, exhibits
a distinct asymmetry with a strong rise towards $\cos\theta^*=-1$, {\it i.e.},
the photon direction. 
This behaviour suggests that events with $x_{\rm obs}^\gamma<0.75$ are
dominantly produced by charm quarks coming from the photon side.
On the other hand, the $\cos\theta^*$ distribution for {\it direct-enriched}
events, with $x_{\rm obs}^\gamma>0.75$, is almost symmetric, as expected for
subprocesses like the one depicted in Fig.~\ref{fig1}(a).
In order to substantiate these observations from the theoretical side, we now
investigate the differential cross section $d\sigma/d\cos\theta^*$ of
$ep\to D^{*\pm}j+X$ in photoproduction for $p_T^D>3$~GeV, $|\eta^D|<1.5$,
$p_T^j>6$~GeV, and $-1.5<\eta^j<2.4$ at NLO.
In Fig.~\ref{cos}, this cross section is decomposed in two ways: in the
physically well-defined direct- (dotted blue histogram) and resolved-enriched
(solid red histogram) contributions, with $x_{\rm obs}^\gamma>0.75$ and
$x_{\rm obs}^\gamma<0.75$, respectively; and in the mathematically defined
direct- (dashed pink histogram) and resolved-photon (dot-dashed green
histogram) contributions.
From Fig.~\ref{cos}, we observe that the direct- and resolved-enriched
contributions, which can be measured experimentally, exhibit very similar
$\cos\theta^*$ dependences as their theoretical counterparts, which, taken
separately, do not represent physical observables.
In other words, the enriched contributions possess a rather high {\it purity}.
Furthermore, we can confirm the findings of Ref.~\cite{Chekanov:2003bu}
concerning the $\cos\theta^*$ dependences of the direct- and resolved-enriched
samples: the former is almost symmetric, whereas the latter is exhibits a
steep rise towards the photon direction.
This demonstrates that the bulk of the resolved-photon contribution is due to
charm component, in accordance with our conclusions from
Figs.~\ref{figlug0}(a) and (b).

\section{Conclusions}

Using results from Refs.~\cite{Fontannaz:2002nu,ephox}, we evaluated the cross
section of $D^{*\pm}$ plus jet associated photoproduction at HERA to NLO in
the parton model of QCD implemented in the ZMVFNS with non-perturbative FFs
extracted from LEP1 \cite{bkkc}, and studied various distributions of it.
As the most important result, we found that preliminary ZEUS data \cite{zeus}
are nicely described by our theoretical predictions.
This may be partly attributed to the fact that the acceptance cut
$p_T^j>6$~GeV \cite{zeus} excludes events at energy scales of order $m_c$ and
below, which cannot be reliably described in this theoretical framework.
In order to obtain a reliable prediction in the low-$p_T$ range as well, the
finite-$m_c$ effects must be included, by working in the GMVFNS.
In the case of $D^{*\pm}$ inclusive photoproduction, this was recently done
for the direct-photon contribution in Ref.~\cite{ks3}.
The good agreement with the ZEUS data \cite{zeus} provides successful tests of
the universality and the scaling violations of the FFs, which are predicted by
the factorization theorem and the DGLAP \cite{dglap} evolution, respectively.

Unfortunately, the variation of the NLO predictions due to scale changes is
larger than the one stemming from using different contemporary NLO sets of
photon PDFs, so that the latter cannot be further constrained by measurements
of $D^{*\pm}$ plus jet associated photoproduction at HERA.

As for the relative importance of the various partons inside the resolved
photon, charm was found to greatly dominate, while the gluon turned out to be 
practically irrelevant.
In particular, the dominance of the charm component in the resolved photon
manifests itself in the characteristic $\cos\theta^*$ dependence of the cross
section, a feature that was already exploited in experimental analyses
\cite{Chekanov:2003bu}.
The comparison of the ZEUS data \cite{zeus} with our NLO predictions 
establishes the presence of intrinsic charm inside the resolved photon with
overwhelming significance and thus supports the validity and usefulness of the
ZMVFNS endowed with non-perturbative FFs in the kinematic regime considered.

\begin{acknowledgments}
We would like to thank E.~Gallo, L.~Gladilin, J.H.~Loizides, and F.~Sefkow for 
helpful discussions about
Refs.~\cite{Breitweg:1998yt,Chekanov:2003bu,kohno,zeus,h1} and J.H.~Loizides
for making available to us the preliminary ZEUS data \cite{zeus} in numerical
form.
B.A.K. thanks the Max-Planck-Institut f\"ur Physik for the hospitality
extended to him during a visit when this work was finalized.
This work  was supported by the Bundesministerium f\"ur Bildung und Forschung
through Grant No.\ 05~HT4GUA/4.
\end{acknowledgments}

\end{document}